\documentclass[journal,10pt]{IEEEtran}

\IEEEoverridecommandlockouts
\usepackage{cite}
\usepackage[pdftex]{graphicx}
\usepackage{caption}
\usepackage{epstopdf}
\usepackage{dblfloatfix}
\usepackage{blindtext}
\usepackage{float}
\usepackage[cmex10]{amsmath}
\usepackage{algorithmic}
\usepackage{amssymb}
\usepackage{color}
\usepackage{bm}
\usepackage{amsfonts}
\usepackage{amsmath}
\usepackage{amsmath,mathtools,amssymb}
\DeclareMathOperator{\sign}{sign}
\DeclarePairedDelimiter{\norm}{\lVert}{\rVert}
\usepackage[mathlines]{lineno}

\newcommand{\cc}[1]{#1}

\begin{document}

\title{Compressed CSI Feedback With Learned Measurement Matrix for mmWave Massive MIMO}

\author{Pengxia~Wu,
	    Zichuan~Liu,~and
        Julian~Cheng,~\IEEEmembership{Senior~Member,~IEEE}

\thanks{P. Wu and J. Cheng are with the School of Engineering, The University
of British Columbia, Kelowna, BC V1X 1V7, Canada (e-mail:pengxia.wu@ubc.ca, julian.cheng@ubc.ca).}
\thanks{Z. Liu is with the School of Electrical and Electronic Engineering, Nanyang Technological University, Singapore, 639798 Singapore.}
}
\maketitle

\begin{abstract}    
 
    A major challenge to implement the compressed sensing method for channel state information (CSI) acquisition lies in the design of a well-performed measurement matrix to reduce the dimension of sparse channel vectors. The widely adopted randomized measurement matrices drawn from Gaussian or Bernoulli distribution are not optimal. To tackle this problem, we propose a fully data-driven approach to optimize the measurement matrix for beamspace channel compression, and this method trains a mathematically interpretable autoencoder constructed according to the iterative solution of sparse recovery. The obtained measurement matrix can achieve near perfect CSI recovery with fewer measurements, thus the feedback overhead can be substantially reduced.

\end{abstract}

\IEEEpeerreviewmaketitle

\section{Introduction}    
    
    Compressed sensing (CS) technique provides a promising alternative for channel state information (CSI) acquisition in \cc{milimeter wave (mmWave)} massive multiple-input multiple-output (MIMO) systems\cite{choi2017compressed,gao2018compressive}. 
	The main idea of these CS based channel acquisition approaches	\cite{choi2015downlink,alkhateeby2015compressed,gao2017reliable,eltayeb2014compressive}
	 is to exploit the beamspace sparsity and formulate the channel estimation problem into a sparse recovery task. 
	 It is well known that the measurement matrix plays an essential role in sparse recovery\cite{alkhateeby2015compressed,heath2016an}. However, due to simplicity, most of existing works use random matrices as measurement matrices. 
	 
	 Unfortunately, the widely adopted randomized measurement matrices drawn from Gaussian or Bernoulli distribution are not optimal for all channel realizations.
	Although it has been shown that several random measurement matrices can achieve perfect recovery with high probability when the dimension of compressed measurements is sufficiently large, random matrices often perform unsatisfactorily in practical applications especially when the dimension of compressed measurements is insufficient\cite{choi2017compressed}. 
	Since the dimension of compressed measurements determines the size of training and feedback overhead, it is meaningful to reduce the number of measurements under the accuracy constraint of sparse recovery.
	
	Compared with random matrices, the deterministic measurement matrix is more appealing because it requires fewer measurements\cite{lotfi2018a}, but the design of deterministic measurement matrix lacks guidelines. Moreover, the deterministic measurement matrices designed in an ad hoc manner do not perform well for different channel realizations.
	Therefore, our goal is to search for an effective method to generate a \mbox{well-performed} measurement matrix that can be used for all channel realizations.
	
    A good measurement matrix can be constructed by exploiting the data features\cite{arora2018compressed, wu2018sparse}. 
    Many real-world datasets have embedding features that can be exploited to perform dimension reduction operations. However, it is yet known whether additional features beyond sparsity exist in mmWave massive MIMO channels. 
    Fortunately, owing to state-of-the-art deep learning (DL) technology, the hidden data features can be effectively learned by neural networks.
    Because our goal is to construct a measurement matrix that performs a linear transformation on beamspace channel vectors, the conventional black-box DL architecture with non-linear operations are unsuitable for our problem.
       
%

\cc{
In this letter, we introduce $l_1$-minimization autoencoder ($l_1$-AE)\cite{wu2018sparse} and propose a \mbox{data-driven} compressed CSI feedback scheme for mmWave massive MIMO systems. 
Specifically, we employ the $l_1$-AE to learn the measurement matrix from beamspace channel samples; then the learned measurement matrix is applied to perform CS based CSI compression and recovery by conventional linear programming solver.
Unlike the conventional deep learning based methods, which often regard the neural network as a black box to create the end-to-end learning process for CSI acquisition, this work constructs an interpretable autoencoder under the CS framework to perform a data driven dimension reduction for channel vectors.
Moreover, the dimension reduction is achieved by a simple linear transformation, which is easy to implement for the UEs in practical massive MIMO systems.
}

\cc{
Numerical results show that, compared with the random matrices, the learned measurement matrix provides higher recovery accuracy for smaller size channel vectors. 
The proposed $l_1$-AE enhanced CSI feedback scheme can also attain higher achievable rate with lower feedback overhead. 
This result suggests that the beamspace channels have certain underlying features that can be exploited by the neural networks. 
This work demonstrates a useful application of DL techniques for designing mmWave massive MIMO systems.
}

\begin{figure*}[t]
\centering
\normalsize
\includegraphics[width=7in]{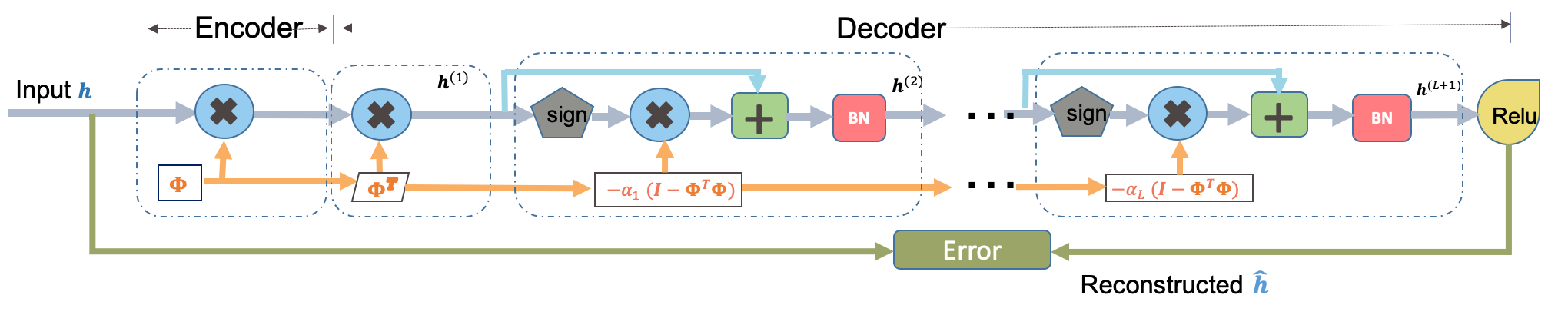}
\captionsetup{justification=centering}
\caption{An $l_1$-AE neural network structure}
\label{L1AE network structure}
\end{figure*}

\section{Beamspace mmWave Massive MIMO Channel}

    We consider a single-cell downlink mmWave massive MIMO system operating in frequency division duplexing (FDD) mode, where a base station (BS) is equipped with $N$ antennas and all user equipments (UEs) are equipped with single antenna. 
    The channel vector for the $k$th user is given by\cite{brady2013beamspace}
\vspace{-1mm}    
\begin{IEEEeqnarray*}{lCl}\label{spatial channel vector}
\cc{\bm h_k^*} = \sqrt{\frac{N}{P}} \sum _{i=1} ^P \beta_k^{(i)} \bm \alpha (\phi _k ^{(i)}) \IEEEyesnumber
\end{IEEEeqnarray*}
    where $P$ is the number of paths; $i = 1$ is the index for the line-of-sight path; $2 \leq i \leq P$ is the index for non-line-of-sight paths; $\beta_k^{(i)}$ is the complex path gain;
    $\phi _k ^{(i)}$ denotes the spatial direction of the $i$th path, and $\bm \alpha (\phi _k ^{(i)})$ is the corresponding array steering vector that contains a list of complex spatial sinusoids. 
    The spatial direction $\phi_k ^{(i)}$ relates to the physical angle $\theta_k ^{(i)}$ by $\phi_k ^{(i)} = \frac{d}{\lambda} \sin{\theta_k ^{(i)}}$, for $-1/2 \leq \phi_k ^{(i)} \leq 1/2$, $-\pi/2 \leq \theta_k ^{(i)} \leq \pi/2$\cite{brady2013beamspace}, where $\lambda$ is the wavelength of mmWave, and $d = \lambda /2$ is the antenna spacing. 
The array steering vector is $\bm \alpha (\phi_k ^{(i)}) = \frac{1}{\sqrt{N}} [e^{-j2\pi \phi_k ^{(i)} (-\frac{N-1}{2})},  e^{-j2\pi \phi_k ^{(i)} (1-\frac{N-1}{2})}, ..., e^{-j2\pi \phi_k ^{(i)} (\frac{N-1}{2})}]^T$ for uniform linear array with $N$ antennas.


The spatial channel vector $\cc{\bm h_k^*}$ in \eqref{spatial channel vector} can be transformed into the beamspace channel representations $\widetilde{\bm h}_k$ by \cite{brady2013beamspace}
\begin{IEEEeqnarray*}{lCl} \label{beamspace channel vector}
 \widetilde{\bm h}_k = \mathbf{U} \cc{\bm h_k^*}  \IEEEyesnumber
\end{IEEEeqnarray*}    
where $\mathbf U$ is the discrete fourier transform matrix of size $N \times N$, and it can be expressed as $\mathbf{U}  = [\bm \alpha(\overline{\phi}_1),  \bm \alpha(\overline{\phi}_2),  ...  , \bm \alpha(\overline{\phi}_N)]^H$,
    where $\overline{\phi}_m = \frac{1}{N} (m-\frac{N+1}{2})$ for $m=1,2,...,N$ is the spatial direction predefined by antenna array, and $\bm \alpha(\overline{\phi}_m)$ is the array steering vector.
The beamspace sparsity is an important feature for mmWave massive MIMO channels.
    The limited number of multipaths $P$ also indicates the limited number of spatial directions $\phi _k^{(i)}$ in \eqref{spatial channel vector}, which corresponds to a fact that a small number of non-zero elements exist in beamspace channel vector $\widetilde{\bm h}_k$\cite{bajwa2010compressed}.

\vspace{-2mm}
\section{$l_1$-AE enhanced compressed CSI feedback}
    
    \cc{This work focuses on the CSI feedback task under the assumption that downlink channel estimation has been completed and feedback links are perfect. 
    The task is to feedback the obtained beamspace channel vector \mbox{$\widetilde{\bm h} \in \mathbb{C} ^ {N \times 1}$} for one UE; therefore, without loss of generalization, we omit the subscript of $\widetilde{\bm h}_k$.} 
    Because the neural network only works with real numbers, we convert the complex channel vector $\widetilde{\bm h} \in \mathbb{C} ^ {N \times 1}$ to the corresponding \mbox{real-valued} compressive channel vector $\bm h \in \mathbb{R} ^ {2N \times 1}$ by stacking its real part on its complex part.  
    
    The compressed measurement vector $\bm y$ is obtained by $\bm y = \mathbf{\Phi} \bm{\bm h}$, where $\bm y \in \mathbb{R} ^{m \times 1}$, and where $m$ denotes the dimension of compressed measurements;
    $\bm \Phi \in \mathbb{R} ^{m \times 2N}$ is the measurement matrix where $m \ll 2N$. 
    In the compressed CSI feedback scheme, the UE sends the compressed measurement vector $\bm y$ with much reduced dimension to the BS; the BS reconstructs the compressive channel vector $\bm h$ based on the received compressed measurement vector $\bm y$ and the measurement matrix $\bm \Phi$. 
    The required number of feedback parameters is therefore reduced from $2N$ to $m$, and the value of $m$ determines the feedback overhead. We desire to make $m$ as small as possible while guaranteeing the recovery accuracy.
    The reconstruction performance highly depends on the measurement matrix $\bm \Phi$, which projects the high-dimensional vector $\bm h$ onto the lower-dimensional subspace spanned by the columns of $\bm \Phi$.
    A good measurement matrix should be designed by taking into account of underlying structural information of the beamspace channel vectors. 
    In order to extract the additional hidden features beyond sparsity in beamspace channels, we propose the $l_1$-AE to learn directly based on beamspace channel vector samples.

\vspace{-2mm}
\subsection{$l_1$-AE Neural Network}

    As shown in \mbox{Fig. \ref{L1AE network structure}}, the $l_1$-AE neural network contains a linear encoder and a dedicated multi-layer non-linear decoder, which are jointly trained to minimize the difference between the input $\bm h$ and the output $\hat{\bm h}$. 
\cc{The $l_1$-AE is built by emulating a complete CS framework, so its structure is interpretable. 
Specifically, the encoder performs the linear compression; the decoder reconstructs $\hat{\bm h}$ by unfolding the iterative solution of sparse recovery, so that the multiple layers of the decoder perform the iterative steps of  recovery algorithms.
More importantly, the $l_1$-AE regards the whole process of  compression and reconstruction as a set of stacked neural networks parameterized with the measurement matrix. 
Therefore, by backpropagating the reconstruction error through the network, the measurement matrix is optimized based on training dataset.}

\emph{Compressive sensing linear encoder:}
    The encoder of \mbox{$l_1$-AE} is simply a matrix-vector multiplication $\bm y  = \bm \Phi \bm h$, where the dimension of the sparse channel vector $\bm h$ is reduced by the measurement matrix $\bm \Phi$; the dimensional-reduced measurement vector $\bm y$ is the output of the encoder, and it is also the input of the decoder. 

\emph{Projection subgradient descent unfolded decoder:}
    The decoder is designed to reconstruct the sparse channel vector $\bm h$. The idea is to unfold the projection subgradient descent algorithm of the $l_1$-minimization optimization for sparse recovery, and each update of the iteration is unfolded as one layer of the decoder.
    The sparse recovery problem is formulated into an $l_1$-minimization optimization problem as
\begin{IEEEeqnarray*}{lCl}\label{norm1 minimization}
\mathop {\min }\limits_{\bm h} \norm{\bm h}_1 \qquad  \text{s.t.} \quad \bm \Phi \bm h=\bm y \IEEEyesnumber 
\end{IEEEeqnarray*}
where $\norm{\bm h}_1$ represents the $l_1$-norm of vector $\bm h$. 
The projection subgradient update of the $l_1$-minimization optimization in \eqref{norm1 minimization} is given by \cite{boyd2003subgradient}
\begin{IEEEeqnarray*}{lCl}\label{projected gradient method}
\bm h^{(t+1)} = \mathcal{P}[\bm h^{t} - \alpha _t \cdot \sign (\bm h^{t})]  \IEEEyesnumber 
\end{IEEEeqnarray*}
where $t$ indicates the $t$th update; $\alpha _t$ is the step size; $\sign(\bm{h}^{t})$ is the subgradient of $\norm{\bm{h}^{t}}_1$; $\mathcal{P}$ indicates the projection onto the convex set \mbox{$\{ \bm h: \bm \Phi \bm h = \bm y \}$}. 
This projection operation on a given vector $\bm x$ is defined as
\begin{IEEEeqnarray*}{lCl} \label{projection}
\mathcal{P}[\bm x] \triangleq \bm x + \bm \Phi ^\dagger (\bm y - \bm \Phi \bm x) \IEEEyesnumber
\end{IEEEeqnarray*}
where $\bm \Phi ^\dagger  = \bm \Phi ^T (\bm{\Phi} \bm{\Phi}^ T)^{-1}$ is the pseudoinverse of $\bm \Phi$. 
    
    According to the projection subgradient descent in \eqref{projected gradient method}, we can obtain the $t$th-step update $\bm{h}^{(t+1)}$ by substituting $\bm{x} = \bm{h}^{t} - \alpha _t \cdot \sign(\bm{h}^{t})$ into \eqref{projection} and set the step size as $\alpha _t = \frac{\alpha}{t}$.
In this way, the $t$th ($1 \le t \le L$) layer decoder can be expressed as
\begin{IEEEeqnarray*}{lCl}\label{decoder update}
\bm h^{(t+1)} = \bm h^{t} - \frac{\alpha}{t} (\mathbf{I} - \bm{\Phi}^T \bm{\Phi}) \sign(\bm h^ t).\IEEEyesnumber 
\end{IEEEeqnarray*}
It is worth mentioning that the pseudoinverse of $\bm{\Phi}^\dagger$ in \eqref{projection} can be replaced by the simple transpose operation $\bm{\Phi}^T$ without performance degradation\cite{wu2018sparse}, so that the computation of back propagation can be simplified.      
    The first layer of decoder is set to be $\bm h^{(1)} = \bm{\Phi}^T \bm{y}$.  
    Additionally, each layer is added by a batch normalization (BN) module to empirically enhance the neural network performance.

    The output layer adopts a \cc{Rectified Linear Unit (ReLU)} activation function, so the reconstructed channel vector $\hat{\bm h}$ is
\begin{IEEEeqnarray*}{lCl}\label{reconstruct output}
\hat{\bm h} &=& \text{ReLU}(\bm h^{(L+1)}) \IEEEyesnumber 
\end{IEEEeqnarray*}
where $\text{ReLU}(\bm h^{(L+1)})$ denotes that for each element $h_i, 0 \le i \le 2N-1$ of $\bm h^{(L+1)}$, the operation $\max\{h_i, 0\}$ is performed.

\emph{\cc{Loss function for training} :} The loss function is defined as the mean square $l_2$-norm error between $\bm{h}$ and $\hat{\bm h}$ samples
\begin{IEEEeqnarray*}{lCl}\label{MSE}
loss = \frac{1}{n} \sum _{i=1}^{n} \norm{\bm{h}-\hat{\bm h}}_2 ^2 \cc{\IEEEyesnumber}
\end{IEEEeqnarray*}
where $n$ is the number of training samples.

\emph{Computational Complexity:}
    The network complexity of $l_1$-AE is mainly associate with computing the weight matrices $\mathbf{I} - \bm{\Phi}^T \bm{\Phi}$ from the \mbox{second-layer} decoder to the \mbox{$(L+1)$th-layer} decoder. Thus, the complexity of \mbox{$l_1$-AE} is about $O(mN^2 L)$. 
    Note that for the structured weight matrix $\mathbf{I} - \bm{\Phi}^T \bm{\Phi}$, the number of independent parameters is only $2mN$. This design of structured weight matrix reduces computation complexity significantly. Because a fully-connected layer requires $2N\times2N$ independent parameters in the weight matrix, which is much more computationally complex when $N$ is large. 
    
It is worth pointing out that the training of $l_1$-AE is conducted offline. Moreover, the offline training is only required once. Hence the training of $l_1$-AE does not consume additional time or spectrum resource of the communication system.

\subsection{$l_1$-AE Enhanced Compressed CSI Feedback}       
    Once the training of $l_1$-AE is completed, a learned measurement matrix $\bm \Phi^*$ as the optimized weight parameters can be extracted from the trained $l_1$-AE network.
    Then the learned measurement matrix $\bm \Phi^*$ is applied to perform the compressed CSI feedback scheme. 
    The process of $l_1$-AE enhanced compressed CSI feedback can be described in three steps.
    First, the training process is performed at the BS, which has large computation power and large data set. The BS shares the learned measurement matrix $\bm \Phi^*$ with its UEs. 
    Second, each UE uses $\bm \Phi^*$ to compress its beamspace channel vectors by the simple multiplication $\bm y = \bm \Phi^* \bm{h}$. The compressed channel vector $\bm y$ is sent to the BS. 
    Third, based on the knowledge of measurement matrix $\bm \Phi^*$ and the feedback vector $\bm y$, the sparse channel vector $\bm h$ can be recovered by a sparse recovery algorithm at the BS.

\section{Experiments and Results}

\subsection{Experiment and Training Parameters}
    We consider a massive MIMO system with 256 antennas for the BS and single antenna for the UE. 
    The channel vector samples are generated according to the channel model in \eqref{spatial channel vector}, and the number of paths is set to be three. 
We randomly generate $20,000$ channel vector samples and then split them into training, development, and test dataset by the ratio of $0.8/0.1/0.1$. 
    Stochastic gradient descent (SGD) method is used to train the $l_1$-AE, and the training parameters are set as follows: learning rate is $0.01$; batch size is $128$; the maximum number of epochs is $1,000$. The measurement matrix $\bm \Phi$ is initialized by the truncated normal distribution with standard deviation $\sigma = 1/\sqrt{512}$. The number of decoder layers is 10, i.e. $L = 9$; the step size $\alpha$ is initialized as $\alpha = 1.0$, and the value of $\alpha$ will be automatically updated to an appropriate value during training. 
 
    We pre-process data to adapt to the valid input-output range of neural network by scaling and shifting the nonzero entries to the [0, 1] range for all the samples. Thus, the original data formation can be easily recovered by performing the corresponding inverse process on the outputs.
    The training takes $2-10$ minutes using a desktop computer equipped with 3.2GHz Intel Core i7-8700 CPU for a given dimension $m$. After that, we obtain the learned measurement matrix $\mathbf{\Phi^*}$.

\subsection{Analysis of Experimental Results}
To assess the performance of the learned measurement matrix $\mathbf \Phi^*$, we compare it with five baseline schemes, which are random Gaussian matrix $\mathbf{G}$, random Bernoulli matrix $\mathbf B$, partial Fourier matrix $\mathbf F$, random selection matrix $\mathbf S$ \footnote{For random selection matrix, entries are $0$ or $1$ with equal probability\cite{choi2017compressed}.}, and random phase shifter matrix $\mathbf P$\footnote{For random phase shifter matrix, each entry is in the form of $e^{j\xi}$, where $\xi$ is randomly selected from a set of quantized angles\cite{alkhateeby2015compressed}.}. 
We use linear programming to perform sparse recovery. The recovery performance is evaluated over the test dataset. 

\begin{table}[]
\centering
\caption {Exact recovery percentages of sparse recoveries with different measurement matrices} \label{exact_percent} 
\begin{tabular}{|p{0.7cm}|l|l|l|l|l|l|l|}
\hline
 Matrix  & $m=20$  & $m=25$  & $m=30$  & $m=35$  & $m=40$    \\ \hline
$\mathbf \Phi^*$  & $95.90\%$ & $98.70\%$ & $99.60\%$ & $100\%$   & $100\%$    \\ \hline
$\mathbf F$          & $0.90\%$  & $7.85\%$  & $89.20\%$ & $99.80\%$ & $99.75\%$  \\ \hline
$\mathbf S$         & $5.30\%$  & $30.15\%$ & $72.70\%$ & $90.00\%$ & $98.45\%$  \\ \hline
$\mathbf B$         & $5.90\%$  & $26.80\%$ & $63.10\%$ & $87.70\%$ & $99.10\%$   \\ \hline
$\mathbf G$        & $2.15\%$  & $13.45\%$ & $58.50\%$ & $84.75\%$  & $97.70\%$    \\ \hline
$\mathbf P$        & $0.00\%$  & $0.00\%$  & $0.45\%$  & $1.00\%$  & $6.85\%$   \\ \hline
\end{tabular}
\end{table}

\begin{figure}[!t]
\centering
\includegraphics[width=3.8in]{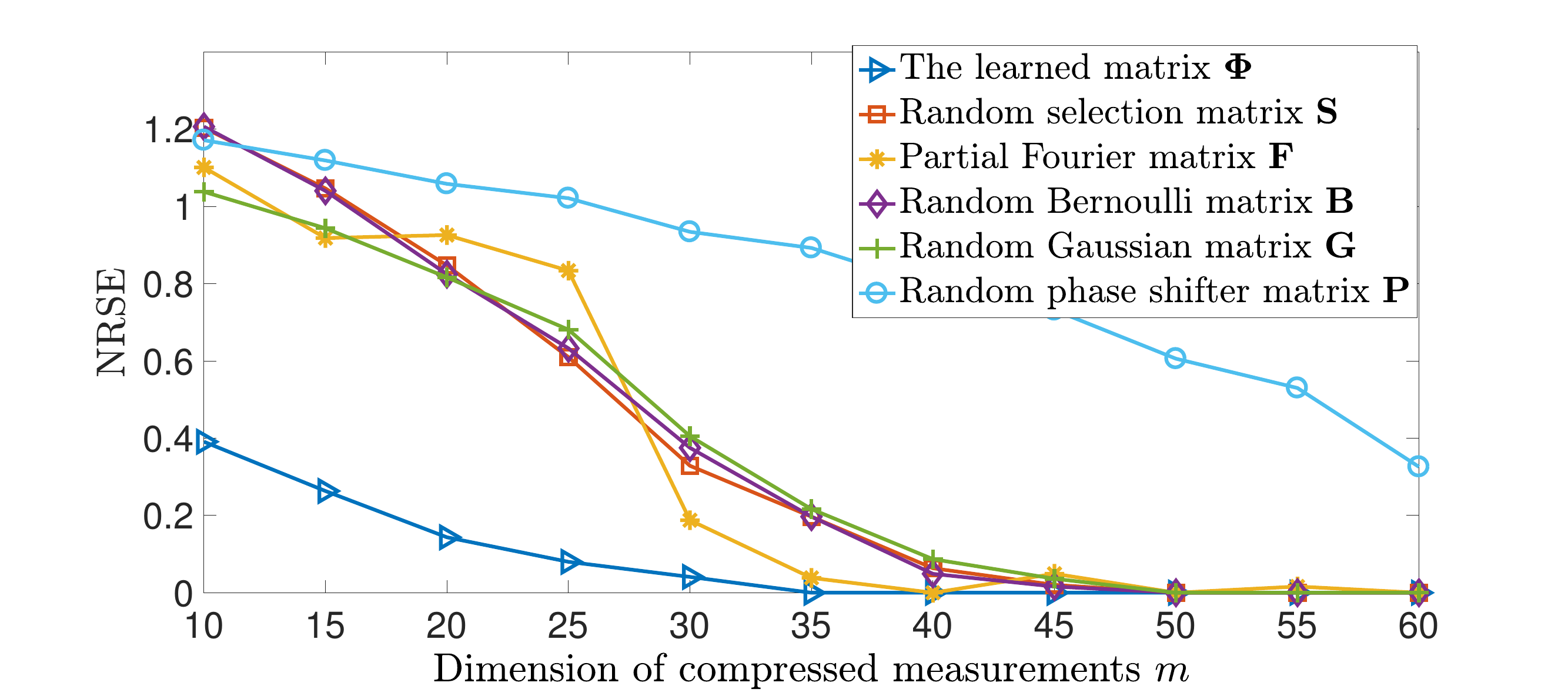}
\caption{Normalized root square error (NRSE) of sparse recovery using different measurement matrices}
\label{NRSE}
\end{figure}

\begin{figure}[!t]
\centering
\includegraphics[width=3.8in]{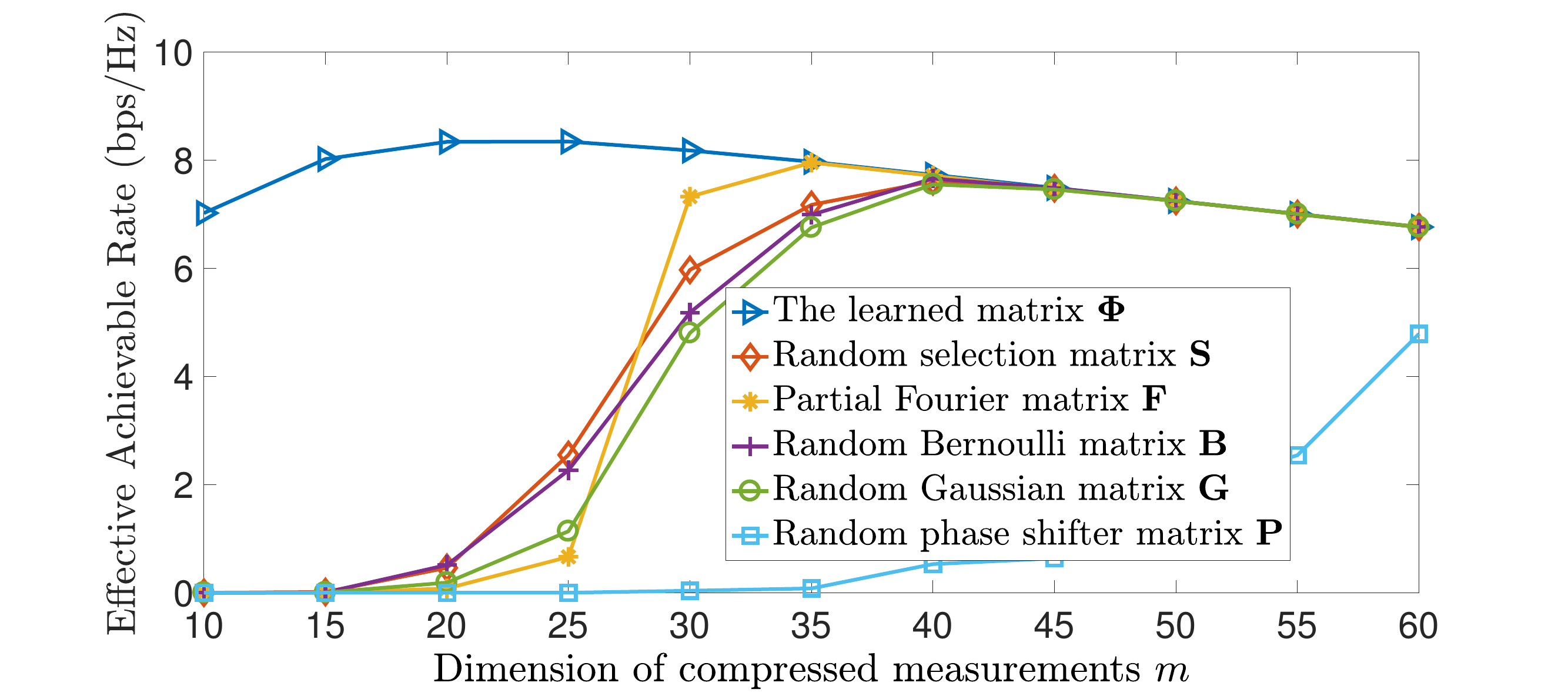}
\caption{Effective achievable rate of feedback CSI compression using different measurement matrices}
\label{sum rate}
\end{figure}

    Table \ref{exact_percent} shows the exact recovery percentages over the test dataset for different measurement matrices, where one sample is counted as recovery if $\norm {{\bm h} - \hat{\bm h}}_2 \le 10^{-8}$. When $m=20$, the learned matrix $\mathbf \Phi^*$ achieves $95.9\%$ recovery, whereas for random matrices the recovery percentages are all less than $6\%$. When the learned matrix $\mathbf \Phi^*$ achieves $98.7\%$ recovery percentage at $m=25$, the highest recovery percentage of random matrix is only $30.15\%$ for random selection matrix $\mathbf{S}$. When $m\ge35$, the $\mathbf \Phi^*$ can achieve perfect ($100\%$) recovery, while none of the random matrices can achieve the same performance.

    Figure \ref{NRSE} compares the normalized root square error (NRSE) of sparse recoveries.
    The learned matrix $\mathbf \Phi^*$ achieves the lowest NRSE for the same dimension of measurements when compared with random matrices.
    In other words, the learned matrix $\mathbf \Phi^*$ can achieve the same level of recovery accuracy with fewer measurements when compared with random matrices. 
  
    A larger dimension of compressed measurements $\bm y$ will lead to better recovery, but lower spectrum efficiency. In order to analyze the trade-off between the number of measurements $m$ and the recovery accuracy, following \cite{alkhateeby2015compressed}, we define the effective achievable rate as $R_{e} = R_0 (1 - \frac{m}{B}) p$, where $R_{0}$ is the maximal achievable rate for one user, $\frac{m}{B}$ is the pilot occupation ratio of one transmission block, $B$ is the block length which is set as 200 symbols, and $p$ is the probability of successful recoveries.
    As shown in Fig. \ref{sum rate}, the effective achievable rate attains maximum at $m=20$ when using the learned matrix $\mathbf \Phi^*$, while for random matrices $\mathbf{S, B, G}$ the maximum effective achievable rates are achieved at $m=35$ or $m = 40$. Moreover, the maximal effective achievable rate for the learned matrix $\mathbf \Phi^*$ is higher than those using random matrices.
    The tremendous performance gain obtained by the learned measurement $\mathbf \Phi^*$ over the random matrices suggests that the sparse beamspace channels have underlying structural features that can be exploited by the DL technique. 

\vspace{-3mm}
\section{Conclusion}
    We proposed a \mbox{data-driven} compressed CSI feedback approach for downlink CSI acquisition of FDD systems. In such a scheme, a fully \mbox{data-driven} measurement matrix was constructed by the \mbox{$l_1$-AE} to enhance the CS method. 
    Compared with the conventional CS methods using random projections, the proposed \mbox{$l_1$-AE} can exploit the hidden data structures of beamspace channel datasets, hence the channel vectors can be compressed into smaller size at the UE and can still be recovered almost perfectly at the BS.
    \cc{As a future research topic, we will study the design of measurement matrix for quantized feedback vectors.}
\vspace{-3mm}

\bibliographystyle{IEEEtran}
\bibliography{IEEEabrv,mybib}

\end{document}